\documentclass[onecolumn]{aastex62}

\graphicspath{{./}{figures/}}
\usepackage[caption=false]{subfig}
\usepackage{natbib}

\shorttitle{Lekshmi, Nandy \& Antia}

\begin{document}

\title{Asymmetry in Solar Torsional Oscillation and the Sunspot Cycle}

\correspondingauthor{Dibyendu Nandy}
\email{dnandi@iiserkol.ac.in}

\author{Lekshmi B}
\affiliation{Center of Excellence in Space Sciences India, Indian Institute of Science Education and Research Kolkata, Mohanpur 741246, West Bengal,India}
\author{Dibyendu Nandy}
\affiliation{Center of Excellence in Space Sciences India,  Indian Institute of Science Education and Research Kolkata, Mohanpur 741246, West Bengal,India}
\affiliation{Department of Physical Sciences,  Indian Institute of Science Education and Research Kolkata, Mohanpur 741246, West Bengal,India}
\author{H.M.Antia}
\affiliation{Tata Institute of Fundamental Research, Homi Bhabha Road, Mumbai 400005, India}



\begin{abstract}

Solar torsional oscillations are migrating bands of slower- and faster-than-average rotation, which are strongly related to the Sun’s magnetic cycle. We perform a long-term study (16 yr) of hemispherical asymmetry in solar torsional oscillation velocity using helioseismic data for the first time. We study the north-south asymmetry in the
velocity using the zonal flow velocities obtained by ring diagram analysis of the Global Oscillation Network Group (GONG) Doppler images. We find significant hemispherical asymmetry in the torsional oscillation velocity and explore its variation with respect to depth, time, and latitude. We also calculate the hemispherical asymmetry in the surface velocity measurements from the Mount Wilson Observatory and the zonal flow velocities obtained from the Helioseismic and Magnetic Imager ring diagram pipeline. These asymmetries are found to be consistent with the asymmetry obtained from GONG observations. We show that the asymmetry in near-surface torsional oscillation velocity is correlated with the asymmetry in magnetic flux and sunspot number at the solar surface, with the velocity asymmetry preceding the flux and sunspot number asymmetries. We speculate that the asymmetry in torsional oscillation velocity may help in predicting the hemispherical asymmetry in sunspot cycles.

\end{abstract}


\section{Introduction} \label{sec:intro}

Solar torsional oscillations are bands of slower- and faster- than-average rotation that migrate from midlatitudes toward the equator and poles during the solar cycle. They were first discovered by \cite{1980ApJ...239L..33H} using full-disk velocity measurements from the Mount Wilson Observatory 150 ft tower. Migrating zonal flow bands were clearly observed using Mount Wilson Doppler observations from 1986 \citep{1998IAUS..185...59U,2001ApJ...560..466U}. \citet{1997ApJ...482L.207K} were the first to observe the flow helioseismically using the Michelson Doppler Imager (MDI) on board the \textit{Solar and Heliospheric Observatory} \citep{1995SoPh..162..129S}, and the migration was observed by \cite{1999ApJ...523L.181S}. The migrating flows were also observed using the helioseismic measurements from  the Global Oscillation Network Group (GONG; \cite{1996Sci...272.1284H}). The first radially resolved evidence of zonal flow migration using GONG was reported by \cite{2000SoPh..192..427H}, and \cite{2000SoPh..192..437T} reported similar findings based on MDI observations. \cite{2000ApJ...533L.163H} concluded that the equatorward band of torsional oscillation penetrates through the convection zone. Observations by \cite{2000ApJ...541..442A} also revealed that they are not just surface phenomena but have depth dependence. \cite{2001ApJ...559L..67A} also studied the poleward propagation of the high-latitude branch of torsional oscillation, which has higher amplitude compared to the equatorward branch. The equatorward branch of the torsional oscillations is strongly related to the Sun's magnetic cycle, with the center of the magnetic activity belt located at the poleward boundary of the fast zone. It has been proposed that the equatorward drift of the oscillation pattern can be produced by Lorentz force due to migrating dynamo-generated fields \citep{1981A&A....94L..17S}, thermal feedback \citep{2003SoPh..213....1S}, or magnetic quenching \citep{1999A&A...343..977K, 1999A&A...346..295P}, or can be due to all the three effects combined \citep{2007ApJ...655..651R}.

Global helioseismology, which uses globally coherent modes of oscillations to study large-scale flows on the Sun, cannot be used to investigate the asymmetry in zonal and meridional flows of the Sun since the splittings of global modes are not sensitive to asymmetry across the equator. The asymmetries in the flows can be studied using local helioseismology techniques \citep{1988ApJ...333..996H, 1993Natur.362..430D, 1997ApJ...485..895L}. Ring diagram analysis \citep{1988ApJ...333..996H, 2000JApA...21..353B}, which is a local helioseismology technique, considers the displacement of three-dimensional power spectra on small areas of solar disk to infer the zonal and meridional flows. We can study the north-south asymmetry using these zonal flow velocities obtained from ring diagram analysis. A drawback of the ring diagram technique is that it is constrained only to the near-surface layers because of the short-wavelength waves that are considered for the analysis. The information on the time-dependent radial and latitudinal gradient of torsional oscillation can be helpful in deducing the temporal variation of the toroidal magnetic field \citep{2008ApJ...681..680A}. Torsional oscillation can also provide information on the timing of the solar cycle before the appearance of new cycle magnetic activity \citep{2011JPhCS.271a2074H}. This information, along with the hemispherical variation of torsional oscillation, could be a prior indicator of the solar cycle. 

\cite{2003ApJ...585..553B} studied the asymmetry in rotation for nine carrington rotations by applying ring diagram analysis on MDI data and concluded that the antisymmetric component becomes significant above 0.98R$_\odot$ and errors are large below this radius to infer any asymmetry. \cite{2006SoPh..236..227Z} also studied the north-south asymmetry in zonal flow for the declining phase of cycle 23 (2001 -- 2004) using the velocities obtained by the ring diagram analysis of GONG Doppler measurements. They investigated the temporal, latitudinal, and radial variation of zonal flow for both hemispheres and its variation with respect to the magnetic cycle. \cite{2011JPhCS.271a2074H} and \cite*{2014SoPh..289.3435K}, using ring diagram results from GONG, reported the hemispherical asymmetry in zonal flow and proposed that the flow pattern can be a precursor of magnetic activity. \cite{2014ApJ...789L...7Z} and \cite{2016AdSpR..58.1457Z}, using time-distance analysis of Helioseismic and Magnetic Imager (HMI) Doppler measurements, also reported the existence of hemispherical asymmetry in torsional oscillation.

 \cite{1983IAUS..102..101H} used 16 yr of Mount Wilson full-disk velocity measurements and \cite{1997SoPh..170..389J} used rotation data obtained from sunspot tracers to study the temporal variation in the north-south asymmetries of solar differential rotation. \cite{2005SoPh..227...27G}, using the rotation of H$\alpha$ filaments, confirmed the existence of north-south asymmetry in solar rotation and proposed a relation with asymmetry in solar activity. 

The aim of this work is to look into the long-term variation in the north-south asymmetry of solar torsional oscillations. We use the zonal flow velocities obtained from the ring diagram analysis of GONG Dopplergrams for a period of 16 yr (2001 July to 2017 March), which are made available from the GONG ring diagram analysis pipeline. We study the variation in the hemispherical asymmetry of torsional oscillations with time, latitude, and depth. We also calculate the velocity asymmetry near the solar surface considering surface velocity measurements from the Mount Wilson Observatory made using the Babcock magnetograph. We compare this with the asymmetry at the near-surface layer obtained from GONG observations. The asymmetry calculated from the zonal flow velocities obtained from the ring diagram pipeline of Helioseismic and Magnetic Imager (HMI) onboard \textit{Solar Dynamics Observatory (SDO)} are also used for comparison. We find these to be qualitatively consistent. The migration of zonal flow and  active region magnetic flux are strongly related. Motivated by this, we perform a correlation analysis of the hemispherical asymmetry in torsional oscillation with the asymmetry in sunspot flux and number. We discover that the asymmetry in velocity precedes the asymmetries in the sunspot cycle and there exists a significant correlation between the two.
\section{Data}

\begin{figure}[ht]
\subfloat[]{
 \includegraphics[width=\linewidth]{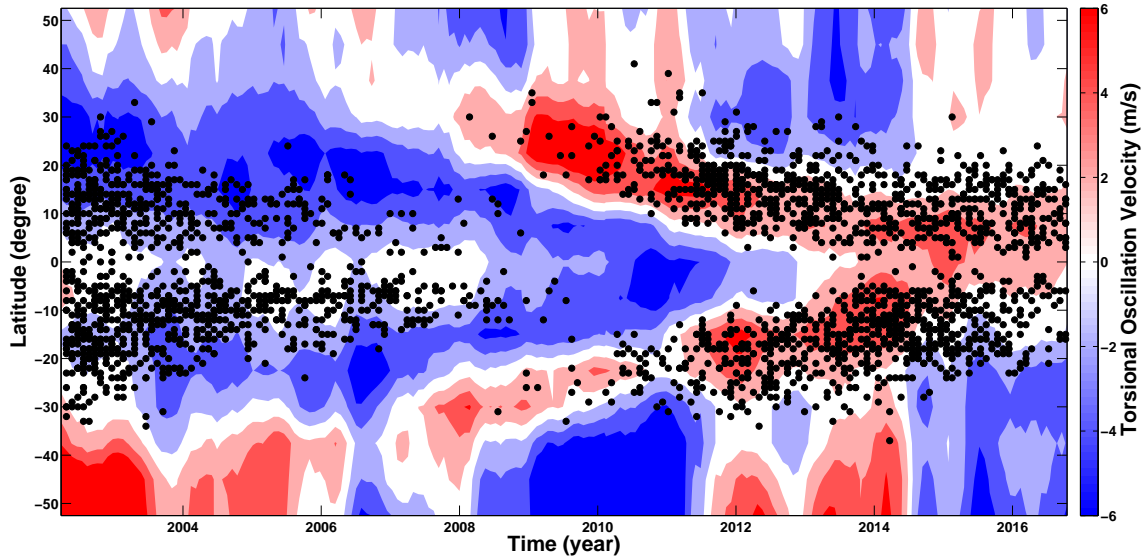}
}

\subfloat[]{
  \includegraphics[width=\linewidth]{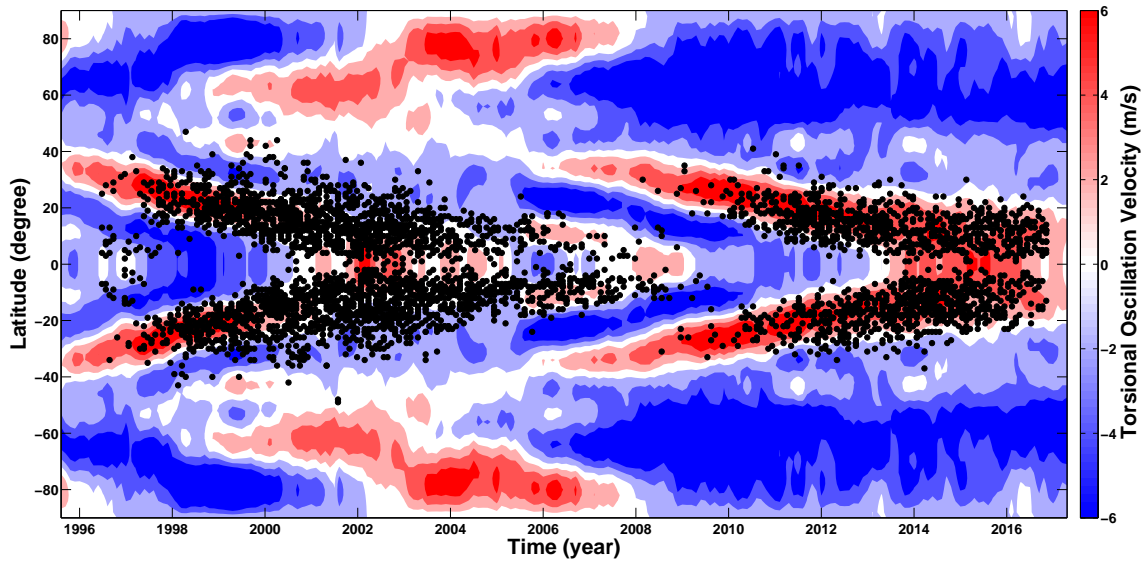}
}

\caption{(a) Contour plot of torsional oscillation (smoothed over 1 yr) at a depth of 2 Mm from solar surface obtained from GONG ring diagram analysis as a function of latitude and time overplotted with sunspot distribution (black circles). The zonal flow velocities are corrected for the effects due to $B_{0}$-angle variation. A 11 year average is subtracted from the velocities at each latitude. The data cover the declining phase of cycle 23 and cycle 24 until 2016. (b) Contour plot of torsional oscillation velocity at a depth of 2 Mm from the solar surface obtained by the global mode inversion of GONG data for the period covering the declining phase of cycle 23 and rising phase of cycle 24. Black circles represent the sunspot distribution for this period.}
\end{figure}

We use the zonal flow velocity data obtained from GONG ring diagram analysis pipeline for the period 2001 July to 2017 March. Ring diagram analysis is performed using the dense pack technique as described by \cite{2002ApJ...570..855H}. The daily Dopplergrams from different GONG sites are merged and remapped into 189 overlapping patches. Each patch is circularly apodized with a diameter of $15^{\circ}$ with centers separated by $7^{\circ}.5$ in latitude and longitude. The latitudinal extent and the longitudinal extent from  central meridian are $-52^{\circ}.5$ to $52^{\circ}.5$ respectively. Each patch is tracked at the Snodgrass rate for 1664 minutes to remove the differential rotation effect. Three-dimensional power spectra are calculated for each patch, and these form rings in the ($k_{x}, k_{y}$) plane. These rings get distorted in the presence of flows, and they are fitted and inverted to obtain the zonal flow and meridional flow velocities as a function of depth. The ring diagram analysis procedure is explained in detail by \cite{2003ESASP.517..255C} and is implemented in the ring diagram pipeline as explained by \cite{2003ESASP.517..295H}. The systematic effects due to the variation in $B_{0}$ angle are corrected, following the procedure mentioned by \cite{2015SoPh..290.1081K}. The data are then smoothed over 1 yr to remove random fluctuations in the data, which remain even after the $B$-angle correction. A 11 yr average over each latitude and depth is subtracted from the zonal flow velocity at the same latitude and depth to obtain the residual rotation rate. Figure 1a is the contour plot of the torsional oscillation velocity as a function of latitude and time, at a depth of 2 Mm from the surface, obtained from GONG observations. Figure 1b is the contour plot of the torsional oscillation velocity obtained by the inversions of global modes obtained from the GONG data \citep{2008ApJ...681..680A}.

The surface velocity measurements obtained from the Mount Wilson 150 ft tower using the Babcock magnetograph \citep{1953ApJ...118..387B} are used for calculating the hemispherical asymmetry at the solar surface and are compared with the near-surface torsional oscillation velocity asymmetry obtained from the GONG observations. The preparation of this data set  is explained by \cite{2001ApJ...560..466U}. The data consist of 34 latitude points with a separation of $4^\circ$ ($-68^{\circ}$ to $68^{\circ}$) and have a temporal resolution of 1.5 days from Carrington rotation 1617 to 2121, with each Carrington rotation divided into 18 intervals. Here we consider the data only for the GONG observing period. For our analysis, we further average the data and interpolate the data for GONG latitudes. The zonal flow derived from the ring diagram analysis of data obtained from HMI is also used for our analysis. $B$-angle correction is applied for these data also, following the procedure suggested by \cite{2015SoPh..290.1081K}, and further smoothed over 1 yr. The data we consider here start from 2010 August and cover a period of about 7 yr with latitude coverage of $-62^{\circ}.5$ to $62^{\circ}.5$. 

The sunspot area data compiled by the Royal Greenwich Observatory (RGO) and the United States Air Force/US National Oceanic and Atmospheric Administration (USAF/NOAA) are used for calculating the sunspot flux used in this study. We consider the maximum area ($A$) of each active region in units of microhemispheres and convert it to magnetic flux by using the empirical relation $\Phi(A) = 7.0 \times 10^{19}\times A$ Mx \citep{1966ApJ...144..723S, 2006GeoRL..33.5102D}. We calculate the magnetic flux for both the hemispheres separately for the period from 2001 July to 2016 September. To check for the relatedness of various proxies for the sunspot cycle, we performed a correlation analysis between this calculated magnetic flux and the flux obtained, from the synoptic magnetograms of MDI and HMI combined together for the same period. A correlation coefficient of 0.88 (99.9$\%$ confidence) is obtained which justifies the use of flux obtained from sunspot proxy for our analysis.

\section{Results}
Asymmetry in torsional oscillation velocity across the equator is clearly observed in the Figure 1a , which is derived from the ring diagram analysis of GONG data. The contour plot of velocity obtained from global modes (Figure 1b), which assumes symmetry across the equator, is included for comparison. Figure 2 is the plot of temporal variation of torsional oscillation velocity for northern (blue) and southern (red) hemispheres at different depths and latitudes obtained from GONG observations. Here we plot the velocities at latitudes $7^{\circ}.5$, $22^{\circ}.5$, $30^{\circ}$, $37^{\circ}.5$ and $45^{\circ}$. This latitude range corresponds to the equatorward migrating branch of torsional oscillation. Local helioseismology restricts the velocity information to shallow depths from the surface. We consider the velocity at radial distances of $0.980$, $0.990$, and $0.997R_\odot$. The time series for the torsional oscillation velocity at $0.990R_\odot$ obtained from global modes at different latitudes is plotted along with the north and south velocities in the respective panels for comparison. Representative error bars (mean) are also shown along with the plots. The error values mentioned here are the mean of standard errors ($\delta$) in velocity values. These are obtained from the standard deviations ($\sigma$) of the velocity values from the mean velocity of each year ($\delta$ = $\sigma/\sqrt{N}$, where $N$ is the number of data points). Our observations cover the declining phase of cycle 23 and rising phase of cycle 24. It is observed that the magnitude of velocity increases with depth in both the hemispheres. Also the velocity amplitude decreases with latitude. It is clear from the figure that an asymmetry exists in velocity between the hemispheres at all depths and latitudes. The velocities appear to be more asymmetric beyond 2011. Also, the asymmetry appears to be high for higher latitudes. We quantify asymmetry as the difference of the velocity in the southern hemisphere from that in the northern hemisphere for each latitude and depth. The comparison with global mode velocity shows that the velocities obtained from both local and global techniques follow the same trend, with local modes providing hemispherical variation and global measurements providing lower errors. Our analysis is restricted above $0.980R_\odot$ because the scatter in the data increases significantly below this depth (the mean standard error of $0.970R_\odot$ is 3.81 m s$^{-1}$ which is two times the error at $0.980R_\odot$ of 1.72 m s$^{-1}$). The standard error is nearly same for the three depths that are used for our analysis.

\begin{figure}[ht]
  \includegraphics[width=\linewidth,scale= 2]{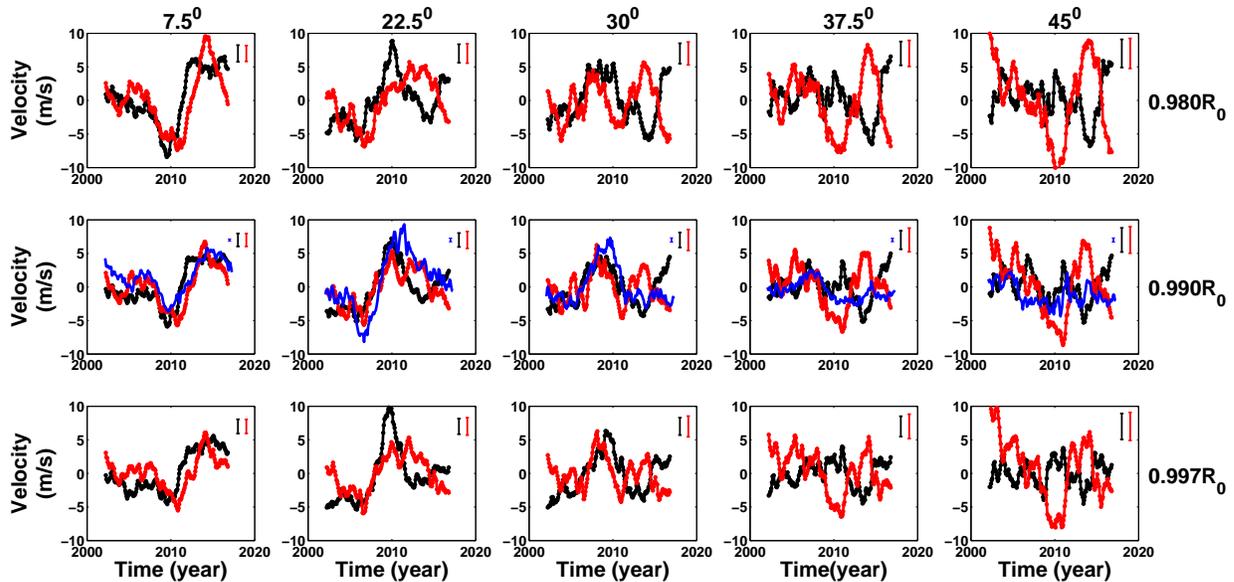}
  \caption{Time series of torsional oscillation velocity in the northern (black) and southern (red) hemispheres at five different latitudes and three different depths. Here we have considered only the velocity of the equatorward-migrating branches for both hemispheres. The zonal flow velocity obtained from the global helioseismic technique at a radial distance of $0.990R_\odot$ at all the five latitudes is plotted in the respective panels (blue). Error bars (mean) for the north (black) and south (red), and global (blue) velocities are also shown.}
\end{figure}

Figure 3 is the plot of hemispherical asymmetry in torsional oscillation velocity (U$_{North}$-U$_{South}$) at different depths and latitudes as a function of time. The positive value of asymmetry implies that the velocity of torsional oscillation in the northern hemisphere is greater than that in the southern hemisphere at the same time, depth, and latitude. It is observed that the asymmetry increases with latitude with almost no asymmetry near solar minimum. When the magnitude in asymmetry is maximum (around 2010 and 2014), it can be observed that there is an increase in asymmetry with depth, though it is not clearly distinguishable in other epochs. The standard error values are comparable to the difference in asymmetry between two consecutive depths, so that we cannot conclusively say whether the observed increase in asymmetry with depth is real or not. Representative error bars (mean) for all depths are given in each panel with the plots. At low latitudes, the time of appearance of maxima shifts to the right with a decrease in latitude for all depths. It can also be clearly seen from Figure 1 that the northern hemisphere branch is migrating faster than its southern hemisphere counterpart for the present cycle. Hence, the northern branch attains maximum velocity at each latitude before the southern hemisphere.

We use the torsional oscillation velocity data prepared from the surface velocity measurements obtained from the Mount Wilson Babcock magnetograph. These data extends from 1974 to 2012. The hemispherical asymmetry in this surface torsional oscillation velocity is calculated for the GONG latitudes. We also calculate the asymmetry for the torsional oscillation velocity obtained from the HMI ring diagram pipeline. For both data sets, we consider the time period overlapping with that of GONG. These asymmetries are compared with the asymmetry obtained from GONG observations for the near-surface layer. Figure 4 is a comparative plot of temporal variation in the asymmetry in velocity obtained from GONG at 2 Mm depth along with the asymmetry of surface velocity obtained from Mount Wilson and asymmetry obtained from HMI data at 2 Mm depth for different latitudes.  Asymmetry from Mount Wilson and HMI observations shows a correlation with that from GONG at low latitudes. For higher latitudes correlation is poor with Mount Wilson but the correlation with HMI is still apparent. The fluctuations in the asymmetry increase with latitude.

\begin{figure}[ht]
  \includegraphics[width=\linewidth]{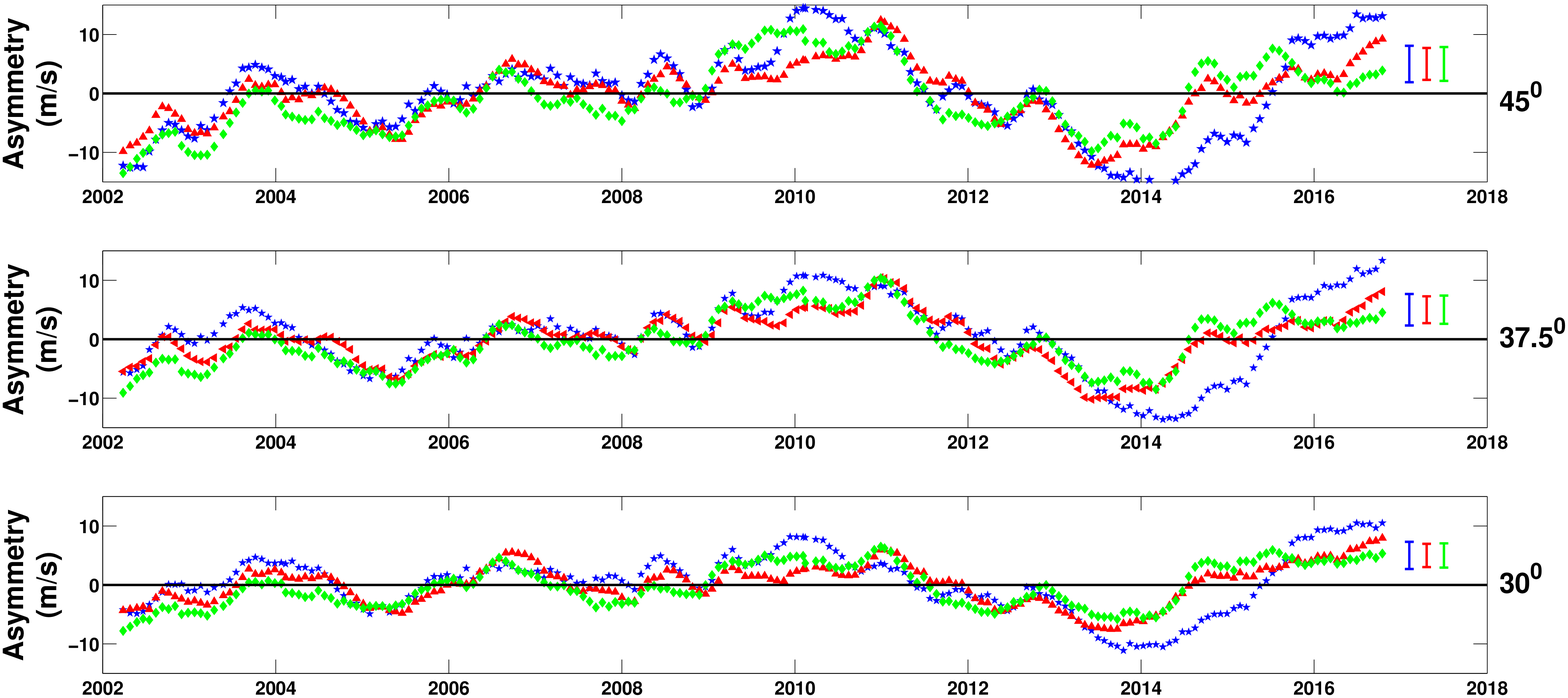}
  \includegraphics[width=\linewidth]{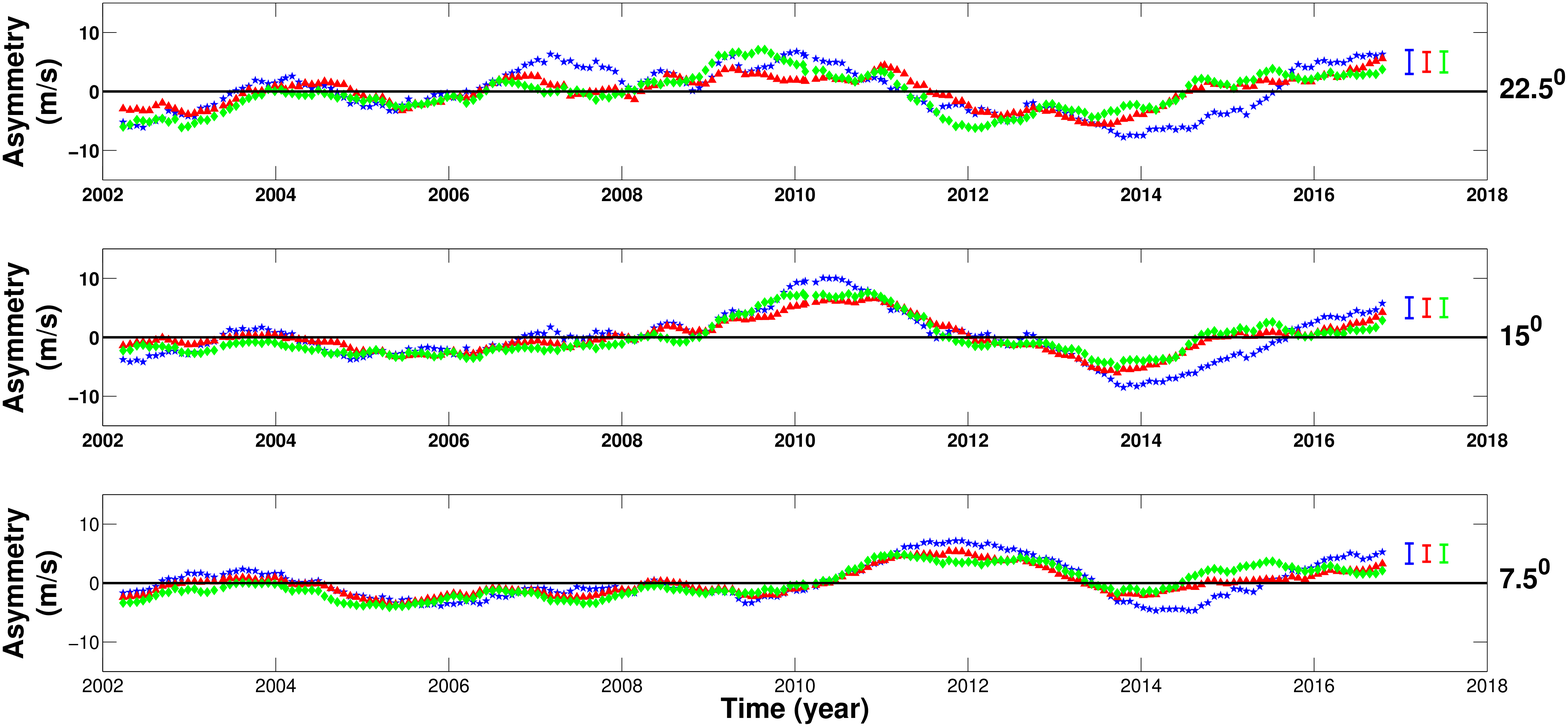}
  \caption{Asymmetry in torsional oscillation velocity ($U_{North}-U_{South}$) for the equatorward-propagating branches for six latitudes at three different depths ($0.980R_\odot$-blue, $0.990R_\odot$-red, $0.997R_\odot$-green). Error bars (mean) corresponding to each depth are shown in respective panels.}
\end{figure}

The solar magnetic cycle and torsional oscillation have similar migrating patterns. Thus, one may expect some correlation between the hemispherical asymmetry in torsional oscillations and magnetic flux. We average the near-surface torsional oscillation velocity (at $0.997R_\odot$) obtained from GONG observations over latitude depending on the location of magnetic features. The data are averaged over $7^{\circ}.5$-$15^{\circ}$ for the time interval 2002.2-2008, $15^{\circ}$-$30^{\circ}$ for 2008-2013 and $7^{\circ}.5$-$22^{\circ}.5$ for 2013-2016.8 in both the hemispheres. The latitude range for each interval is chosen depending on the location of active regions within that interval, such that the averaged velocity encompasses the active latitude. Hemispherical asymmetry is calculated from this averaged velocity over the whole observing period. We calculate the asymmetry in the magnetic flux and sunspot number (SSN) between the hemispheres obtained from RGO and USAF/NOAA sunspot data. Sunspot flux and SSN asymmetries are quantified in the same way as the velocity asymmetry. Figure 5 is the plot of this hemispherical asymmetry in velocity as a function of time, along with the asymmetry in sunspot flux scaled by $0.75 \times 10^{22}$ and SSN for the same period. We observe that the asymmetries in flux and number appear to be correlated with asymmetry in velocity over time with a time delay. The asymmetry in velocity is observed to be preceding the asymmetry in flux and SSN. We generate 1000 random sample sets of velocity asymmetry, flux, and SSN asymmetries based on their mean value and standard error at each time stamp. We do a correlation analysis of this random set of velocity asymmetry with the randomly generated sets of flux and SSN asymmetries for a range of time delays from 0 to 2.25 yr, with the asymmetry in velocity preceding the flux and SSN asymmetries. We obtain 1000 time delay values corresponding to the peak value of the Pearson correlation coefficient for each set. A mean time delay of 1.41 $\pm$ 0.38 yr is obtained with a correlation coefficient of 0.72 (99.9$\%$ confidence) for the correlation between velocity and flux asymmetries. A correlation coefficient of 0.78 (99.9$\%$ confidence) is obtained for a mean time delay of 1.18 $\pm$ 0.23 yr between the velocity asymmetry and SSN asymmetry.


\begin{figure}[ht]
  \includegraphics[width=\linewidth]{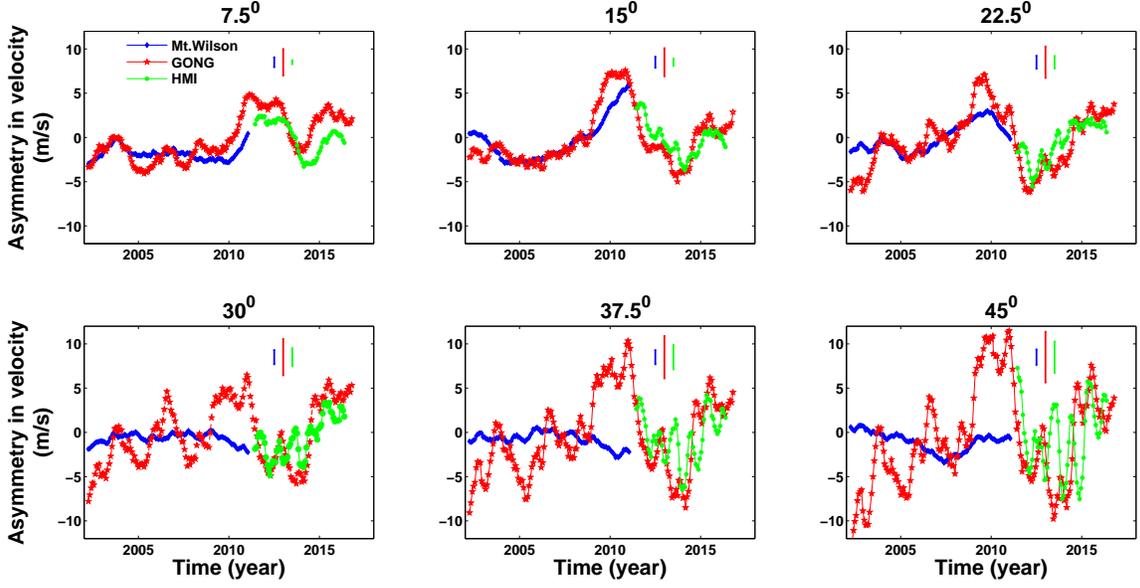}
  \caption{Asymmetry in torsional oscillation at a depth of 2 Mm obtained from the GONG observations (red), Mount Wilson surface velocity measurements (blue), and HMI observations at 2 Mm (green) for different latitudes as a function of time. The asymmetries are considered only for the GONG observing period. Error bars (mean) are given in respective panels.}
\end{figure}

\begin{figure}[ht]
  \includegraphics[width=\linewidth]{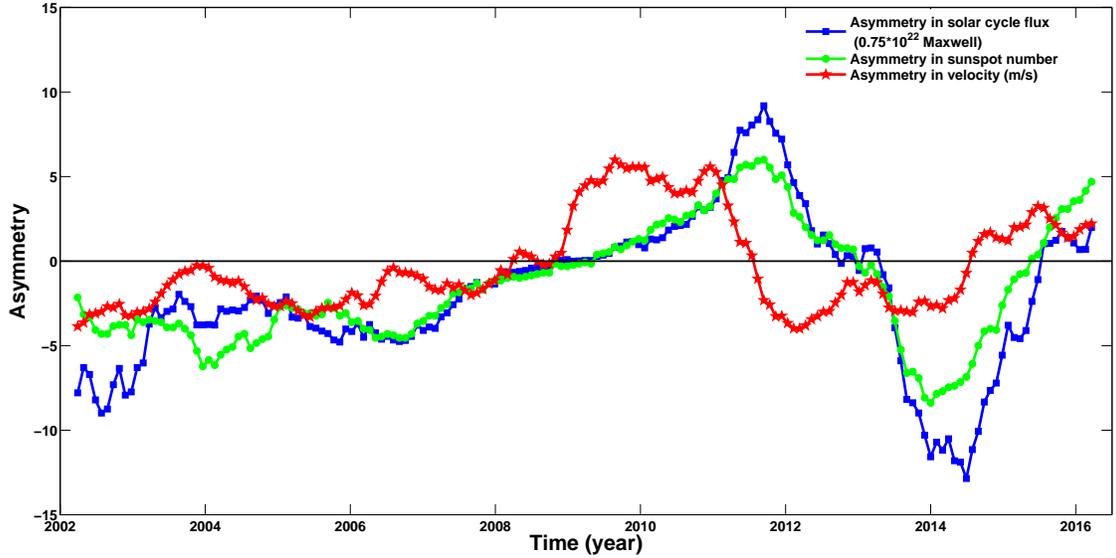}
  \caption{Asymmetry in torsional oscillation (red) at 2 Mm, obtained by taking the latitudinal average of  velocity for both the hemispheres. Latitudinal average is done depending on the position of active regions at different epochs. The asymmetries in sunspot flux (blue) and sunspot number (green) are also plotted as a function of time.}
\end{figure}

Next, we try to relate the asymmetry in velocity with the asymmetry in flux and SSN. The top panel of Figure 6 is a scatter plot of the asymmetry in flux versus asymmetry in torsional oscillation velocity with a time delay of 1.41 yr, with the latter preceding the former. This scatter plot is fitted with a straight line ($p_1x + p_2$) with coefficients $p_1 = 8.47\times10^{21} $Mx s m$^{-1}$ and $p_2 = -6.37\times10^{21}$ Mx within 95$\%$ confidence bounds given by ($7.16\times10^{21}$ Mx s m$^{-1}$, $9.77\times10^{21}$ Mx s m$^{-1}$) and ($-1.00\times 10^{22}$ Mx, $-2.70\times10^{21}$ Mx), respectively. The bottom panel is the scatter plot of asymmetry in sunspot number as a function of asymmetry in torsional oscillation with the velocity asymmetry preceding by 1.18 yr. The coefficients of the fitted line are $p_1 = 0.99$ s m$^{-1}$  and $p_2 = -0.71$ within 95$\%$ confidence bounds given by ($0.87$ s m$^{-1}$, $1.12$ s m$^{-1}$) and $(-1.06, -0.36)$, respectively.

\begin{figure}[ht]
  \includegraphics[width=\linewidth]{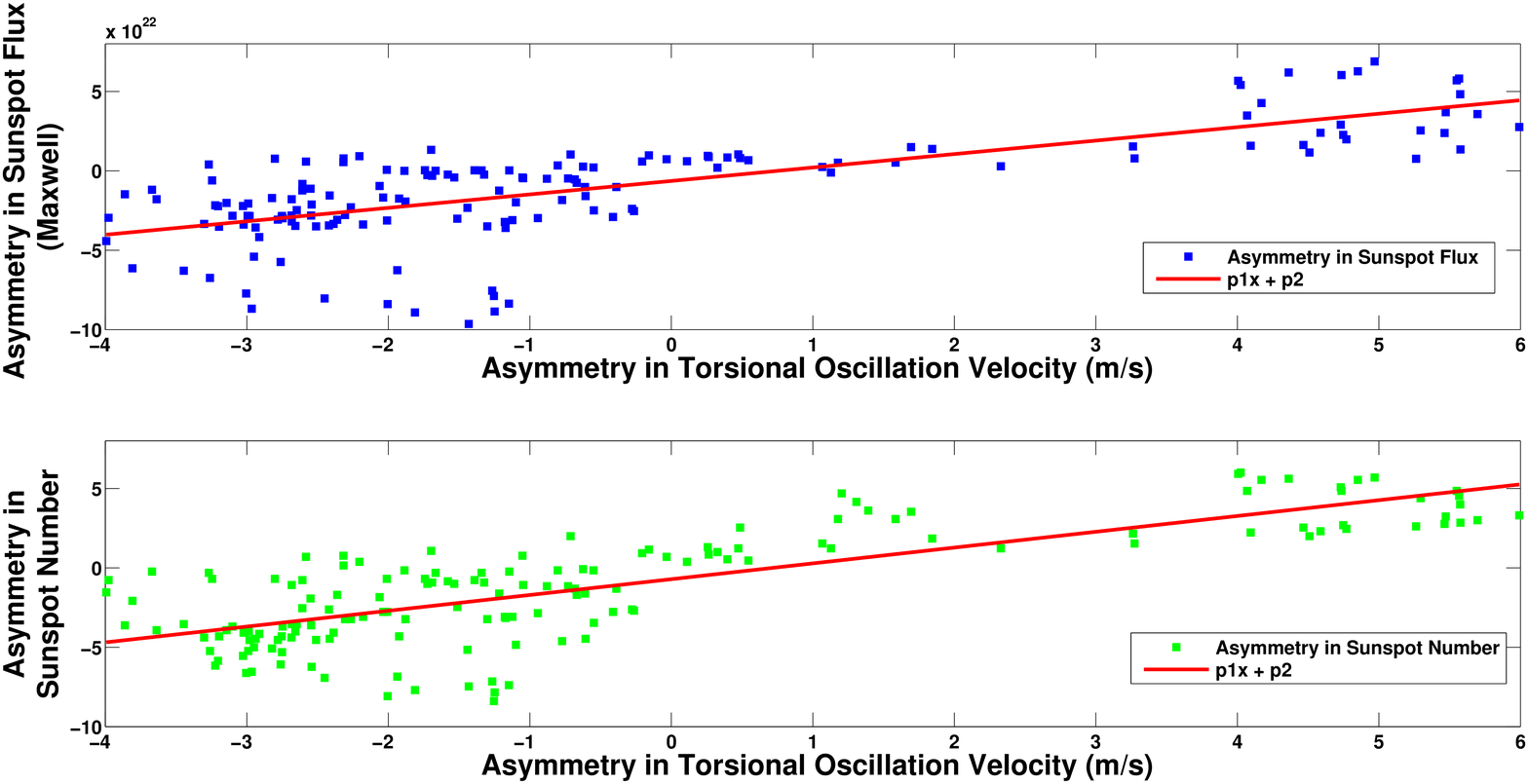}
  \caption{Top panel : Asymmetry in sunspot flux (t yr) versus the asymmetry in torsional oscillation velocity (t-1.41 yr), fitted with a straight line of slope $8.47\times10^{21}$ Mx s m$^{-1}$ and intercept $-6.37\times10^{21}$ Mx. Bottom panel : Asymmetry in sunspot number (t yr) versus the asymmetry in torsional oscillation velocity (t-1.18 yr), fitted with a straight line of slope $0.99$ s m$^{-1}$ and intercept $-0.71$. The mean standard errors in the sunspot flux and number are shown in the respective panels.
  	 }
\end{figure}

\section{Conclusions}
We report an asymmetry between the torsional oscillation velocities of the northern and southern hemispheres at all depths and latitudes. \cite{2003ApJ...585..553B} studied asymmetry in solar rotation for nine Carrington rotations, but any definite temporal variation in the asymmetry component was not observed. However, our study, using data for 16 yr shows that there is temporal variation in the asymmetry and that it tends to be zero near the solar minima. There is an increase in asymmetry beyond 2012 at higher latitudes (Figure 3). A faster-than-average velocity region appeared at high latitudes around 2012-2014 in the southern hemisphere and around 2015-2016 in the northern hemisphere. The increase in asymmetry is because of the delay in the appearance of this region in the northern hemisphere. We speculate that these regions correspond to the poleward-migrating branches of the next sunspot cycle. For lower latitudes, the time of appearance of maximum asymmetry is delayed, and this delay corresponds to the migration of bands from higher to lower latitudes. Also the northern branch is observed to be migrating faster than the southern branch. A similar observation was made by \cite{2014SoPh..289.3435K}. It was observed from the time-distance analysis of HMI data by \cite{2014ApJ...789L...7Z} and \cite{2016AdSpR..58.1457Z} that the faster band of the equatorward-migrating branch of torsional oscillation in the northern hemisphere is closer to the equator than its southern hemisphere counterpart for the present cycle.

 The feedback of magnetic forces on flows can impact the velocities obtained by feature tracking, while these do not directly affect doppler measurements. The difference in the asymmetries obtained from GONG and Mount Wilson velocities at low latitudes can be due to this reason. The absence of magnetic features at higher latitudes can affect the velocity measurements by feature tracking, and hence the Mount Wilson velocities show larger variations from the GONG measurements at these latitudes since the former are based on tracking the migration of magnetic features. The asymmetry derived from GONG observations is well correlated with that derived from HMI observations at all latitudes. Fluctuations and hence errors are high for GONG observations at higher latitudes. The variation between all the three asymmetries at higher latitudes can also be due to projection effects. 

Sunspots appear on the poleward boundary of the faster-rotating branch, and the equator moving bands are associated with the magnetic activity. The northern hemisphere attains peak velocity before the southern hemisphere because of the faster migration of the northern band within this time frame. Interestingly, observations show that the northern hemisphere attains maximum magnetic flux before the southern hemisphere. \cite{1983IAUS..102..101H} combined the torsional oscillation velocities for the northern and southern hemispheres and showed that the torsional oscillation precedes magnetic activity, increasing by $\sim$ 1 yr before the total magnetic flux. We observe that the asymmetry in torsional oscillation appears to precede the asymmetry in magnetic flux by 1.41 $\pm$ 0.38 yr and sunspot number asymmetry by 1.18 $\pm$ 0.23 yr on average and that they are well correlated. We derive empirical relations connecting the asymmetry in velocity with flux and sunspot number. We speculate on the possibility of predicting the hemispherical asymmetry in magnetic cycle using the torsional oscillation asymmetry. \cite{2005SoPh..227...27G} suggest that the asymmetry in solar activity can be responsible for north-south asymmetry in solar differential rotation. But the time delay between the asymmetries, with velocity preceding the magnetic activity, and their strong correlation from our observations point to the possibility that the asymmetry in magnetic flux can be a consequence of asymmetry in torsional oscillations. Alternatively, the asymmetry in the deep-seated magnetic flux belt of a sunspot cycle before they are manifested at the surface may also generate an early asymmetry in near-surface torsional oscillation through a nonlocal (e.g., thermal shadow) effect. Further work with more accurate deep flow diagnostics and magnetohydrodynamic simulations is necessary to conclusively establish the physical process at play. 

This work utilizes data from the National Solar Observatory Integrated Synoptic Program, which is operated by the Association of Universities for Research in Astronomy, under a cooperative agreement with the National Science Foundation and with additional financial support from the National Oceanic and Atmospheric Administration, the National Aeronautics and Space Administration, and the United States Air Force. The GONG network of instruments is hosted by the Big Bear Solar Observatory, High Altitude Observatory, Learmonth Solar Observatory, Udaipur Solar Observatory, Instituto de Astrof$\acute{i}$sica de Canarias, and Cerro Tololo Inter-American Observatory. We are thankful for the courtesy of NASA/\textit{SDO} and HMI science teams. We thank Roger Ulrich for the torsional oscillation data from the Mount Wilson surface velocity measurements. We thank Rachel Howe and Kiran Jain for valuable suggestions and comments regarding the GONG data analysis. We extend our gratitude to Andrew Marble for helping out with the GONG data download. We acknowledge the use of sunspot data from the Royal Greenwich Observatory and the USAF/NOAA active region database compiled by David H. Hathaway. We thank Prantika Bhowmik for providing the sunspot flux data used in our analysis. We are grateful to the Ministry of Human Resource Development, Government of India, for supporting this research at CESSI. D.N. acknowledges the NASA Heliophysics Grand Challenge Grant NNX14AO83G and the Indo-U.S Joint Center Programme grant IUSSTF-JC-011-20 for facilitating interactions that inspired this research.

\end{document}